\documentclass[usenatbib]{mn2e}
\usepackage{graphicx,natbib}
\usepackage{bm,url,color}
\graphicspath{{./fig/}{./png/}}

\newcommand{\EQ}{\begin{equation}}
\newcommand{\EN}{\end{equation}}
\newcommand{\EQA}{\begin{eqnarray}}
\newcommand{\ENA}{\end{eqnarray}}

\newcommand{\Eq}[1]{Eq.~(\ref{#1})}
\newcommand{\Eqs}[2]{Eqs.~(\ref{#1}) and~(\ref{#2})}

\newcommand{\App}[1]{Appendix~\ref{#1}}

\newcommand{\Sec}[1]{Sect.~\ref{#1}}

\newcommand{\Fig}[1]{Fig.~\ref{#1}}

\newcommand{\Figs}[2]{Figs.~\ref{#1} and \ref{#2}}
\newcommand{\Figss}[2]{Figs.~\ref{#1}--\ref{#2}}
\newcommand{\Tab}[1]{Table~\ref{#1}}

{}
{}
{}

{}
{}
{}
{}
{}
{}
{}
{}
{}
{}
{}
{}
{}
{}
{}
{}
{}
\newcommand{\meanUU}{\overline{\bm{u}}}

{}

\newcommand{\meanb}{\overline{b}}

\newcommand{\hatkk}{\hat{\bm{k}}}

{}

{}
{}
{}

\newcommand{\nnn}{\hat{\bm{n}}}

\newcommand{\meanAA}{{\overline{\bm{a}}}}
\newcommand{\meanBB}{{\overline{\bm{b}}}}
\newcommand{\meanbb}{{\overline{\bm{b}}}}
\newcommand{\meanJJ}{{\overline{\bm{j}}}}

\newcommand{\kk}{\bm{k}}

\newcommand{\xx}{\bm{x}}

\newcommand{\rr}{\bm{r}}

\newcommand{\BB}{\bm{B}}

\newcommand{\nab}{{\bm{\nabla}}}

\newcommand{\pom}{\mbox{\boldmath $\varpi$} {}}

\newcommand{\ii}{{\rm i}}
\newcommand{\erf}{{\rm erf}}

\newcommand{\diag}{{\rm diag}  \, {}}

\newcommand{\dd}{{\rm d} {}}

\newcommand{\const}{{\rm const}  {}}

\def\degr{\hbox{$^\circ$}}
\def\la{\mathrel{\mathchoice {\vcenter{\offinterlineskip\halign{\hfil
$\displaystyle##$\hfil\cr<\cr\sim\cr}}}
{\vcenter{\offinterlineskip\halign{\hfil$\textstyle##$\hfil\cr<\cr\sim\cr}}}
{\vcenter{\offinterlineskip\halign{\hfil$\scriptstyle##$\hfil\cr<\cr\sim\cr}}}
{\vcenter{\offinterlineskip\halign{\hfil$\scriptscriptstyle##$\hfil\cr<\cr\sim\cr}}}}}
\def\ga{\mathrel{\mathchoice {\vcenter{\offinterlineskip\halign{\hfil
$\displaystyle##$\hfil\cr>\cr\sim\cr}}}
{\vcenter{\offinterlineskip\halign{\hfil$\textstyle##$\hfil\cr>\cr\sim\cr}}}
{\vcenter{\offinterlineskip\halign{\hfil$\scriptstyle##$\hfil\cr>\cr\sim\cr}}}
{\vcenter{\offinterlineskip\halign{\hfil$\scriptscriptstyle##$\hfil\cr>\cr\sim\cr}}}}}
\def\Rey{\mbox{\rm Re}}
\def\Imag{\mbox{\rm Im}}

\def\urms{u_{\rm rms}}

\def\etaT{\eta_{\rm T}}

\def\Beq{b_{\rm eq}}

\def\half{{\textstyle{1\over2}}}

\newcommand{\mJy}{\,{\rm mJy}}
\newcommand{\beam}{\,{\rm beam}}

\newcommand{\K}{\,{\rm K}}
\newcommand{\g}{\,{\rm g}}
\newcommand{\s}{\,{\rm s}}
\newcommand{\um}{\,\mu{\rm m}}
\newcommand{\uG}{\,\mu{\rm G}}

\newcommand{\cm}{\,{\rm cm}}
\newcommand{\millim}{\,{\rm mm}}

\newcommand{\km}{\,{\rm km}}

\newcommand{\kms}{\,{\rm km\,s}^{-1}}

\newcommand{\pc}{\,{\rm pc}}
\newcommand{\kpc}{\,{\rm kpc}}
\newcommand{\Mpc}{\,{\rm Mpc}}

\newcommand{\Gyr}{\,{\rm Gyr}}

\newcommand{\yaj}[3]{ #1, {AJ,} {#2}, #3}

\newcommand{\yapj}[3]{ #1, {ApJ,} {#2}, #3}

\newcommand{\yapjl}[3]{ #1, {ApJ,} {#2}, #3}
\newcommand{\yapjs}[3]{ #1, {ApJS,} {#2}, #3}

\newcommand{\yan}[3]{ #1, {Astron.\ Nachr.,} {#2}, #3}

\newcommand{\yana}[3]{ #1, {A\&A,} {#2}, #3}
\newcommand{\yanas}[3]{ #1, {A\&AS,} {#2}, #3}

\newcommand{\ysov}[3]{ #1, {Sov.\ Astron.,} {#2}, #3}

\newcommand{\yjetp}[3]{ #1, {Sov.\ Phys.\ JETP,} {#2}, #3}

\newcommand{\yaraa}[3]{ #1, {ARA\&A,} {#2}, #3}

\newcommand{\yprl}[3]{ #1, {PhRvL,} {#2}, #3}

\newcommand{\ymn}[3]{ #1, {MNRAS,} {#2}, #3}
\newcommand{\ynat}[3]{ #1, {Nature,} {#2}, #3}

\newcommand{\yprd}[3]{ #1, {PhRvD,} {#2}, #3}

\newcommand{\yjour}[4]{ #1, {#2}, {#3}, #4}

\newcommand{\ybook}[3]{ #1, {#2} (#3)}
\newcommand{\yproc}[5]{ #1, in {#3}, ed.\ #4 (#5), #2}
\newcommand{\pproc}[5]{ #1, in {#2}, ed.\ #3 (#4), in press, arXiv:#5}

\newcommand{\pana}[2]{ #1, {A\&A}, in press, arXiv:#2}

\title[Helicity proxy for edge-on galaxies]
{Application of a helicity proxy to edge-on galaxies}

\author[Brandenburg \& Furuya]{
Axel Brandenburg$^{1,2,3,4}$\thanks{E-mail:brandenb@nordita.org} and
Ray S. Furuya$^{5}$
\\
$^1$Nordita, KTH Royal Institute of Technology and Stockholm University, Roslagstullsbacken 23, SE-10691 Stockholm, Sweden\\
$^2$Department of Astronomy, AlbaNova University Center, Stockholm University, SE-10691 Stockholm, Sweden\\
$^3$JILA and Laboratory for Atmospheric and Space Physics, University of Colorado, Boulder, CO 80303, USA\\
$^4$McWilliams Center for Cosmology \& Department of Physics, Carnegie Mellon University, Pittsburgh, PA 15213, USA\\
$^5$Institute of Liberal Arts and Sciences, Tokushima University, Minami Jousanajima-machi 1-1, Tokushima 770-8502, Japan
}

\date{\today,~ $ $Revision: 1.139 $ $}
\begin{document}
\maketitle

\begin{abstract}
We study the prospects of detecting magnetic helicity in galaxies by
observing the dust polarization of the edge-on galaxy NGC~891.
Our numerical results of mean-field dynamo calculations show that there
should be a large-scale component of the rotationally invariant parity-odd
$B$ polarization that we predict to be negative in the first and third
quadrants, and positive in the second and fourth quadrants.
The large-scale parity-even $E$ polarization is predicted to be negative
near the axis and positive further away in the outskirts.
These properties are shown to be mostly a consequence of the magnetic
field being azimuthal and the polarized intensity being maximum at the
center of the galaxy and are not a signature of magnetic helicity.
\end{abstract}


\section{Introduction}

The magnetic fields of spiral galaxies possess a clear large-scale
component along with a fluctuating component of comparable strength
\citep{BBMSS96,Han17}.
Owing to the presence of turbulence in the interstellar medium (ISM), there
is significant turbulent diffusion, which would destroy the large scale
magnetic field on a time scale of less than a billion years \citep{Shu98},
unless there is a correspondingly strong anti-diffusive mechanism.
The best known mechanism for explaining the origin and maintenance
of galactic large-scale magnetic fields is the $\alpha$ effect
\citep{Par55,SKR66}.
In the context of galactic dynamo theory, it was first explored by
\cite{Par71} and \cite{VR71}.
The existence of the $\alpha$ effect requires a violation of statistical
mirror symmetry, which implies the presence of magnetic helicity.
It is important to assess the validity of the dynamo models toward
a more comprehensive understanding of how a galaxy forms its structure
and how magnetic fields regulate the cycle of the ISM by enhancing and
suppressing star-formation activity in a galaxy.
However, there is no explicit evidence that the $\alpha$ effect really
does operate in galaxies.

Methods for measuring the helicity of the magnetic field in our Galaxy
and in distant galaxies have been proposed on several occasions in recent
years \citep{VS10,JE11,OJRE11,BS14,HF14}.
Measuring this for edge-on galaxies may be particularly advantageous,
because we can then see both the upper and lower disc planes
simultaneously.
However, to measure the magnetic helicity, one needs the full magnetic
field vector.
Unfortunately, the linear polarization parameters, Stokes $Q$ and
$U$, only allow one to determine the magnetic field direction up to a
$180\degr$ ambiguity.
For this reason it is preferable to work directly with the Stokes
parameters and to determine from them a helicity proxy.
The quantity of interest is then the parity-odd contribution to
the rotationally invariant constituent of the linear polarization,
or possibly the correlation between the parity-even and parity-odd
polarizations \citep{Kamion97,SZ97}.
This is a relatively new and unexplored technique, which has recently been
applied to the Galactic magnetic field \citep[][hereafter BB20]{BB20}
and to the Sun's magnetic field \citep[][hereafter BBKMRPS]{BBKMRPS19};
see also \cite{Bra19} and \cite{Pra+20} for subsequent applications to
the global solar magnetic field and to solar active regions, respectively.
The original motivation came from \cite{KMLK14} in the cosmological
context.

The attributes parity-even and parity-odd mean the same as mirror-symmetric
and mirror-antisymmetric---at least in a statistical sense.
For example, a cyclone on a weather map is not statistically
mirror-symmetric.
In fact, a cyclone in the northern hemisphere looks like a mirror image
of a cyclone in the southern hemisphere.
A cyclone is therefore statistically mirror antisymmetric.
The underlying physical quantity is the kinematic helicity.
It is a pseudoscalar, because it changes sign when the system is viewed
through a mirror or under parity transformation.
The pressure, by contrast, is a scalar; it behaves similarly in the
north and the south in that it decreases in a cyclone, for example.

In dynamo theory, one distinguishes quadrupolar and dipolar magnetic
fields, whose symmetry about the equator is even and odd, respectively.
As a continuous measure of this, one defines the ratio of the difference
of the energies contained in the symmetric and antisymmetric parts,
divided by their sum.
This quantity is sometimes also called parity \citep{BKMMT89}, but it
is not a pseudoscalar and is therefore, in this sense, a misnomer.

Before we define the parity-even and parity-odd constituents of linear
polarization, let us emphasize that these quantities are, at best,
a proxy of magnetic helicity.
As was already demonstrated in BBKMRPS and \cite{Bracco19}, there
cannot be a detailed correspondence with helicity, because polarization
is only defined with respect to a plane.
If the velocity or magnetic fields are statistically isotopic, these
fields can still be fully helical \citep{Mof78}, but a planar image of a
cyclone could, with equal probability, also be seen from its back side,
so it would appear like its mirror image.
Of course, turbulence in galaxies is inhomogeneous, i.e., its statistical
properties are not the same in different places.
An observer can tell whether one is observing from the outside or
the inside.
Therefore, there is a chance (but no guarantee) that a particular
polarization pattern is preferred over its mirror image.
Rotating convection, inspected from the top surface, is such an example
in the context of solar turbulence.

Let us emphasize at this point that the use of the parity-odd constituent
of linear polarization as a proxy of magnetic helicity does not explicitly
require information about the line-of-sight component of the magnetic
field.
Faraday rotation measurements are therefore not invoked in our study.
As was already discussed in \cite{Bra19}, it is not clear
at this point how this could even be done. This can be regarded as a
disadvantage, but we have to keep in mind that our technique is not
trying to reconstruct the magnetic field.
Instead, we use the appropriate and complete description of linear
polarization to obtain a pseudoscalar without the need for a questionable
reconstruction.
It is simply a new diagnostic and it is then up to us to find out whether
or not it has anything to do with magnetic helicity.
It might also have to do with differential rotation, or with a combination
of helicity and differential rotation.

The purpose of the present work is to learn more about the theoretically
expected patterns for edge-on galaxies by taking NGC~891 as a
representative case.
Here, we consider a numerical solution of a simple galactic mean-field
dynamo of $\alpha\Omega$ type.
The magnetic field is generated by the $\alpha$ effect \citep{KR80}
and differential rotation.
We embed a flat dynamo in a Cartesian domain.
We do this by choosing a distribution of $\alpha$ that is concentrated
about the midplane and has opposite signs above and below it.
The magnetic diffusivity is taken to be constant.
The radius of the disc is assumed to be $15\kpc$.
Some of our models are similar to those of \cite{BDMSST93}, who discussed
the application to two particular edge-on galaxies, NGC~891 and NGC~4631.
Recent multi-wavelength continuum emission studies between wavelengths
of $3.6\um$ and $2.6\millim$ clearly show that the properties of dust
grains in the ISM of NGC~891 are similar to those in the Milky Way
\citep{Hughes}.
This makes this galaxy an ideal laboratory for testing not only our model,
but also for studying the roles of magnetic fields in galaxies in general.

At this point, our goal is to identify characteristic and distinguishing
features in the polarization pattern rather than to construct a detailed
prediction for NGC~891, which would include a detailed treatment of the
wind, as was done in \cite{BDMSST93}, and more recently by \cite{MS17}.
The wind can both enhance and suppresses dynamo action \citep{CSS15}.
Most importantly, however, it can make the polarization orientation
more vertical and thus closer to what is observed \citep{EGRW95}.
There is also great interest in modeling the magnetic field in the
galactic halo \citep{SS90,BDMSST92}, where significant magnetic field
strengths are observed \citep{HS99,Wie+15,Kra19,Kra+20}.

\section{Parity-even and parity-odd polarizations}

In the context of polarimetry of the cosmic microwave background
radiation, one commonly decomposes the linear polarization into the
parity-even $E$ and the parity-odd $B$ polarizations.
These $E$ and $B$ fields are defined as the real and imaginary
parts of a certain quantity $R$, which is given in terms of a global
expansion of linear polarization on the full sphere of the sky.
Here, however, we are only interested in local Cartesian patches in
the sky.
We can then employ standard Fourier transformation of the complex
polarization $P=Q+\ii U$ to compute the quantity
\begin{equation}
\tilde{R}(k_x,k_z)=(\hat{k}_x-\ii \hat{k}_z)^2
\int e^{-\ii\kk\cdot\xx} P(\xx) \, \dd^2\xx,
\label{Rk}
\end{equation}
where the tilde indicates Fourier transformation of $R$ over the spatial
coordinates in the projected plane in the sky, which are here the $x$
and $z$ coordinates, so $(x,z)$ is the plane of the sky and $y$
is the line of sight coordinate pointing away from the observer.
The real and imaginary parts of $R$ give the $E$ and $B$ polarizations,
and $\hat{k}_x$ and $\hat{k}_z$ are the $x$ and $z$ components of
the planar unit vector $\hatkk=\kk/k$, with $\kk=(k_x,k_z)$ and
$k=(k_x^2+k_z^2)^{1/2}$ being the length of $\kk$.
Thus, we have
\begin{equation}
E(\xx)+\ii B(\xx)\equiv R=
\int e^{\ii\kk\cdot\xx} \tilde{R}(\kk) \, \dd^2\kk/(2\pi)^2.
\label{EBdef}
\end{equation}
The occurrence of the factor $(\hat{k}_x-\ii \hat{k}_z)^2$ in \Eq{Rk}
is explained by the fact that it is the square of the complex conjugate
of the generating function $\hat{k}_x+\ii \hat{k}_z$ for pure $E$ and
$B$ modes.
Multiplying this by a phase factor $e^{\ii\pi n/4}$ samples $E$
patterns for $n=0$ and $2$ and $B$ patterns for $n=1$ and $3$ in a
continuous fashion \citep{Bra20}; see also standard text books in the
field \citep{Dur08} and reviews \cite{KK16}.

When comparing the signs of $E$ and $B$ with other work in the literature,
one should be aware of the possibility of different sign conventions; see
\cite{Bra19} and \cite{Pra+20} for a more detailed discussion.
We refer to \App{LocGlob} for a comparison relevant to the present work.

The $E$ polarization corresponds to cross-like magnetic field patterns
if $E>0$ and to ring-like magnetic field patterns if $E<0$; see Figure~2
of BBKMRPS.
Positive (negative) $B$ polarizations, on the other hand, correspond
to clockwise (counterclockwise) inward spiraling patterns.
The two latter patterns are evidently parity-odd, because that for $B>0$
becomes the pattern for $B<0$ under parity transformation.
By contrast, the $E$ patterns do not change under parity transformation.
In this paper, we consider an edge-on galaxy, so we are not concerned
with the grand spiral design of galaxies.
This, too, could in principle lead to $B$ polarization (BB20), but this
would require the galaxy to be viewed face-on.

\begin{table}\caption{
Signs of $E$ and $B$ reported in the literature for characteristic
features in the northern hemisphere for various cases and viewing
directions.
}\vspace{12pt}\begin{tabular}{lcccl}
case & $\!\!E\!\!$ & $\!\!B\!\!$ & $\!\!$view$\!\!$ & $\!\!$reference \\
\hline
rotating convection$\!\!$ & $\!\!-\!\!$ & $\!\!-\!\!$ & $\!\!$face-on$\!\!$ & $\!\!$BBKMRPS \\
solar active regions$\!\!$& $\!\!-\!\!$ & $\!\!+\!\!$ & $\!\!$face-on$\!\!$ & $\!\!$\cite{Pra+20} \\
Galaxy                    & $\!\!-\!\!$ & $\!\!-\!\!$ & $\!\!$Sun$\!\!$     & $\!\!$BB20 \\
spherical dynamo          & $\!\!-\!\!$ & $\!\!-\!\!$ & $\!\!$edge-on$\!\!$ & $\!\!$\cite{Bra19} \\
edge-on galaxies          & $\!\!-\!\!$ & $\!\!-\!\!$ & $\!\!$edge-on$\!\!$ & $\!\!$present work \\
\label{Texperience}\end{tabular}\end{table}

Even when the system is inhomogenous and a finite $B$ emerges, it is
not yet clear what its sign is.
This uncertainty may be mainly due to the fact that the $B$ polarization
has not yet been studied under sufficiently many circumstances.
It does depend on the spatial magnetic field pattern produced by the system.
The experience gathered so far is somewhat sketchy; see \Tab{Texperience}
for a summary.
The results depend not only on the nature of the physical system under
consideration, but also on the viewing direction.
Viewing a disk galaxy from the outside, for example edge-on, will produce
the opposite result for the $B$ polarization as viewing from the inside,
for example when viewing our Galaxy from the position of the Sun, as
was done by BB20.

Looking at \Tab{Texperience}, there seems to be agreement regarding
a negative sign for $E$, but $B$ can have either sign, depending on
circumstances.
In rotating convection, $E<0$ and $B<0$; see Fig.~10 of BBKMRPS.
In solar active regions; $E<0$ but $B>0$; see Fig.~5 of \cite{Pra+20}.
Also, within $\pm10\degr$ near the Galactic midplane, $E<0$ and
$B<0$ ($B>0$) for the azimuthally averaged polarization in the northern
(southern) hemisphere; see BB20.
For a spherical mean-field dynamo, Fig.~1 of \cite{Bra19} shows positive
$E$ and $B$ in the north, but this corresponds to $E<0$ and $B<0$.
This would agree with the present paper if we took $E$ near the axis
and $B$ in the first and third quadrants in an edge-on view.
We emphasize again that these properties with respect to north and
south are independent of the question of whether the magnetic field is
even or odd about the equatorial plane.
Determining this with Faraday rotation measurements is certainly of
interest, but it is not explicitly connected with our helicity proxy.

\section{The model}
\label{TheModel}

\begin{figure}\begin{center}
\includegraphics[width=\columnwidth]{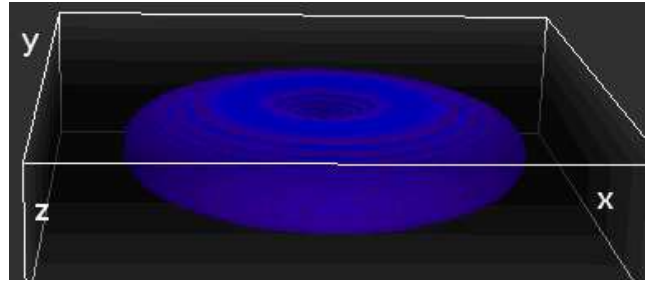}
\end{center}\caption[]{
Three-dimensional visualization of $\BB^2$ for Model~II.
The nearly edge-on observer sees the $xz$ plane and is located in the
negative $y$ direction.
}\label{volviz2}\end{figure}

\begin{figure*}\begin{center}
\includegraphics[width=\textwidth]{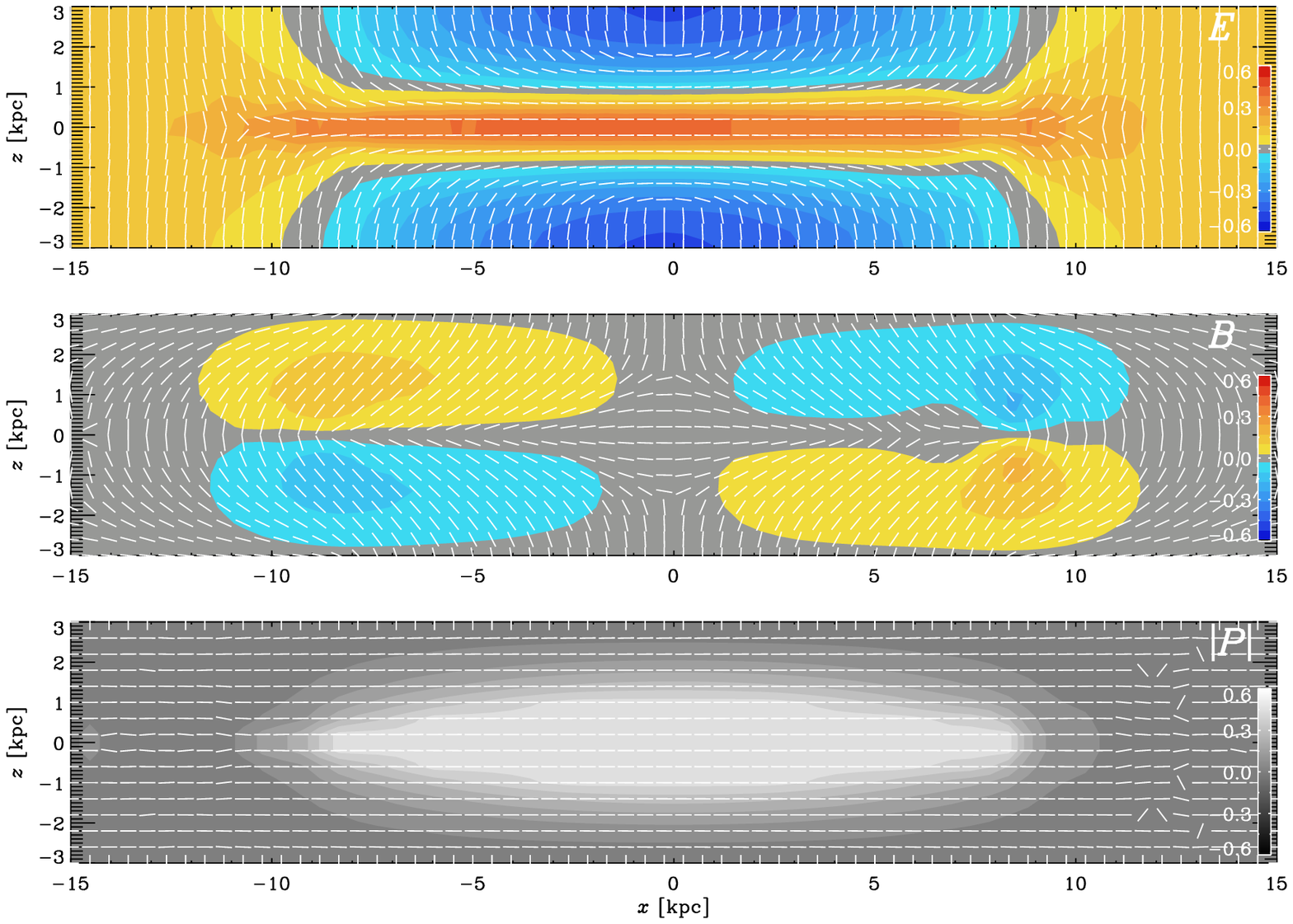}
\end{center}\caption[]{
$E$ and $B$ polarizations as well as polarized intensity $|P|$
in an edge-on view of Model~I, which is the galactic dynamo model
of \cite{Bra15}.
The short lines show the corresponding angles.
The color scale shows the values of $E$, $B$, and $|P|$
in units where $\epsilon=1$ in \Eq{Peqn}; see BBKMRPS.
}\label{pEBxz}\end{figure*}

We adopt the galactic dynamo model of \cite{Bra15}.
It is fully three-dimensional, but when computing an edge-on view,
we consider a specific $xz$ cross-section through $y=-10\kpc$ with the
observer being located in the direction $y\to-\infty$; see \Fig{volviz2}.
Except for one case, where we investigate the magnetic field in the
central slice through $y=0$, we chose the slice $y=-10\kpc$ as a
compromise that is well in front of the central slice and already
sufficiently far into the galaxy to be representative of its field near
the periphery.
We compute the complex polarization, $P=Q+\ii U$, as
\begin{equation}
P=-\epsilon\,(\meanb_x+\ii \meanb_z)^2/\meanbb_\perp^2,
\label{Peqn}
\end{equation}
where $\meanbb_\perp\equiv(\meanb_x,\meanb_z)$ is the mean magnetic
field in the $(x,z)$ plane and $\epsilon$ is the emissivity
\citep{Alton04,PlanckXIX}, which is here assumed to be constant.
The observer is at $y\to-\infty$, which is equivalent with
the definition of BBKMRPS, where the observer of $(b_x,b_y)$ was at
$z\to\infty$, corresponding to the vertical direction in their rotating
convection simulations and their view toward the Sun.
Next, we compute $R=E+\ii B$ using \Eq{EBdef}.
We then show $E(x,z)$ and $B(x,z)$ at a given position $y$; see
\Figss{pEBxz}{pEBxz_thick_pol_128x64_20kc_BDMSST93_wind10a}.
We indicate the polarization angle
\begin{equation}
\chi=\half\mbox{atan}(\Imag P/\Rey P),
\label{chieqn}
\end{equation}
which we also calculate for the pure $E$ and $B$ modes by computing
$\tilde{P}_{E/B}(k_x,k_z)=(\hat{k}_x+\ii \hat{k}_z)^2\tilde{R}_{E/B}(k_x,k_z)$
in Fourier space, where $\tilde{R}_E=\tilde{E}$ and $\tilde{R}_B=\ii\tilde{B}$.

We adopt a mean-field model where all dependent variables are
defined as azimuthal averages indicated by an overbar.
The mean magnetic field $\meanBB$ is expressed in terms of the mean
magnetic vector potential $\meanAA$ as  $\meanBB=\nab\times\meanAA$.
We solve the $\alpha\Omega$ dynamo equation in its simplest form,
\begin{equation}
\frac{\partial\meanAA}{\partial t}=\meanUU\times\meanBB
+\bm{\alpha}\cdot\meanBB-\etaT\mu_0\meanJJ,
\end{equation}
where $\meanUU$ is the mean flow, $\meanJJ=\nab\times\meanBB/\mu_0$
is the mean current density, $\mu_0$ is the vacuum permeability,
$\bm{\alpha}=\diag(\alpha_\perp,\alpha_\perp,0)$ is the $\alpha$ tensor
in the limit of rapid rotation \citep{Rue78}, and $\etaT$ is the total
(microphysical and turbulent) magnetic diffusivity.
The mean flow is composed of toroidal and poloidal components,
$\meanUU_{\rm t}$ and $\meanUU_{\rm p}$, respectively,
describing the galactic rotation profile and a galactic wind.
We use $\meanUU_{\rm p}=W_r\,[1-\exp(-z^2/2H_W^2)]\,\rr/R$
from \cite{BDMSST93} in one case where we include a galactic wind.
The value of $W_r$ determines its strength and $H_W$ is the height
above which the wind commences.

In the following we assume the galactic rotation to be represented
by a modified Brandt rotation profile of the form
$\meanUU_{\rm t}=\pom\Omega(\varpi)$, where $\pom=(x,y,0)$ with
$\varpi=|\pom|$ being the cylindrical radius, and
\begin{equation}
\Omega=\Omega_0/[1+(\varpi/\varpi_\Omega)^n]^{1/n}
\end{equation}
is the angular velocity with $\Omega_0=\const$ characterizing the
rigid rotation law for $\varpi<\varpi_\Omega$, and $\varpi_\Omega=3\kpc$
is the radius where the rotation law attains constant linear velocity
$V_0=\Omega_0\varpi_\Omega$.
The exponent $n$ allows one to make the transition from rigid to constant
rotation sharper, but in the models presented below we used the rather
moderate value $n=2$ for Model~I and $n=3/2$ for Models~II--IV (see
below).

\begin{figure*}\begin{center}
\includegraphics[width=\textwidth]{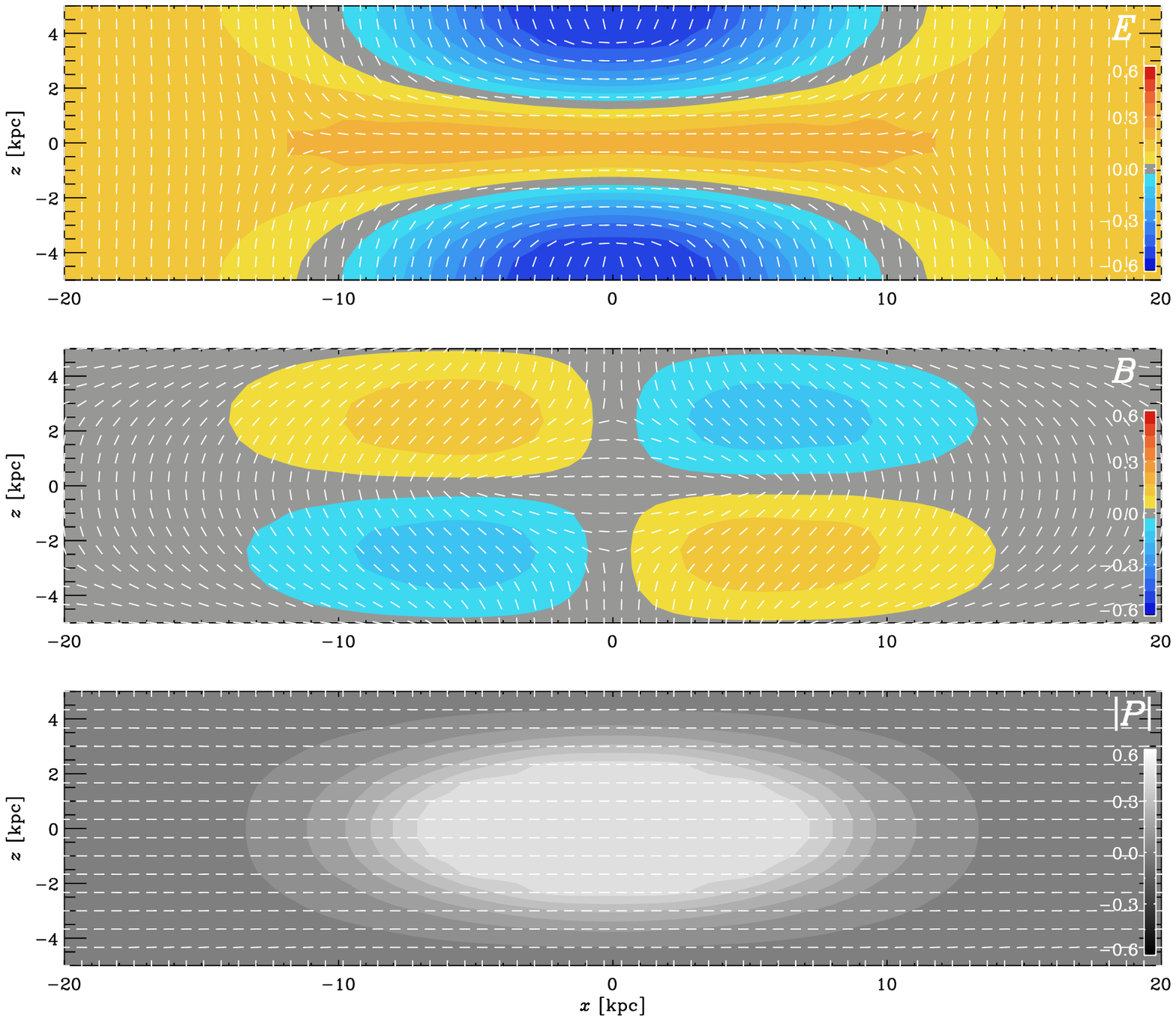}
\end{center}\caption[]{
Same as \Fig{pEBxz}, but for Model~II with parameters relevant to NGC~891.
Note the pixel resolution of about $650\pc$, which corresponds to the
14 arcsecond beam of the JCMT at $850\um$ wavelength.
}\label{pEBxz_64x64_15kc_BDMSST93c_Oml}\end{figure*}

For the $\alpha$ effect we assume a nonlinear ($\alpha$-quenched)
formulation \citep{IR77} for the horizontal components of the
$\alpha$ tensor and choose
\begin{equation}
\alpha_\perp=\frac{\alpha_0}{1+Q_\alpha\meanBB^2/\Beq^2}
\frac{z}{H_\alpha}\exp\left(-{z^2\over H_\alpha^2}\right) f_\alpha(\varpi)\, g_\alpha(\varpi),
\end{equation}
where $H_\alpha$ is the disc height for the $\alpha$ effect,
$\alpha_0$ quantifies its strength, and $f_\alpha(\varpi)$ and
$g_\alpha(\varpi)$ are radial profiles that we apply in Models~II--IV.
In those cases, we use
\begin{equation}
f_\alpha(\varpi)=[1-\erf(\varpi/\varpi_\alpha)]/2
\end{equation}
to introduce a radial cutoff at $\varpi=\varpi_\alpha=15\kpc$, and
\begin{equation}
g_\alpha(\varpi)=[1+(\varpi/\varpi_{\alpha\Omega})^n]^{-1/n}
\end{equation}
to allow for an additional radial modulation with
$\varpi_{\alpha\Omega}=\varpi_\Omega$ in Models~II--IV, such that $\alpha_\perp$
is proportional to the local angular velocity, as was assumed in the
model of \cite{BDMSST93}, who assumed $\alpha_0=\Omega_0\ell$, with
$\ell=0.3\kpc$ being the correlation length.
In Model~I, on the other hand, $\alpha_\perp$ is independent of $\varpi$
and therefore we put $\varpi_{\alpha\Omega}=\infty$.
Note that in both cases, owing to the $z/H_\alpha$ factor in front of the
exponential function, $\alpha_\perp$ changes sign about the midplane.
This reflects the opposite orientation of the Coriolis force on the
two sides of the midplane.
For comparison, we also present a more complicated case with
$W_r=10\kms$ using $H_\alpha=H_W=0.5\kpc$ (Model~III).

We define the equipartition field strength based on the
root-mean-square value of the turbulent velocity, $\urms$, as
$\Beq=\sqrt{\mu_0\rho}\,\urms$; see also Eq.~(1) of \cite{BCEB19}.
Using $\urms=10\km\s^{-1}$ and $\rho=2\times10^{-24}\g\cm^{-3}$, we
have $\Beq=5\uG$ \citep{BDMSST93}, which is the value we use in all
of our models.
For orientation we note that the line-of-sight and plane-of-sky
components of the magnetic field toward galactic molecular clouds
are on the order of $1$--$100\uG$, estimated from Zeeman effect
measurements \citep[e.g.,][]{Crutcher12} and dust polarization studies
\citep[e.g.,][]{PF19}, respectively.
The mean magnetic field is typically below $10\uG$ \citep{BCEB19}.
The parameter $Q_\alpha$ is a nondimensional constant that determines
the strength of $\alpha$-quenching, and thereby the overall magnetic
field strength.
We choose $Q_\alpha=25$, so the resulting mean magnetic field strength
attains plausible values of somewhat below $10\uG$.

\begin{figure*}\begin{center}
\includegraphics[width=\textwidth]{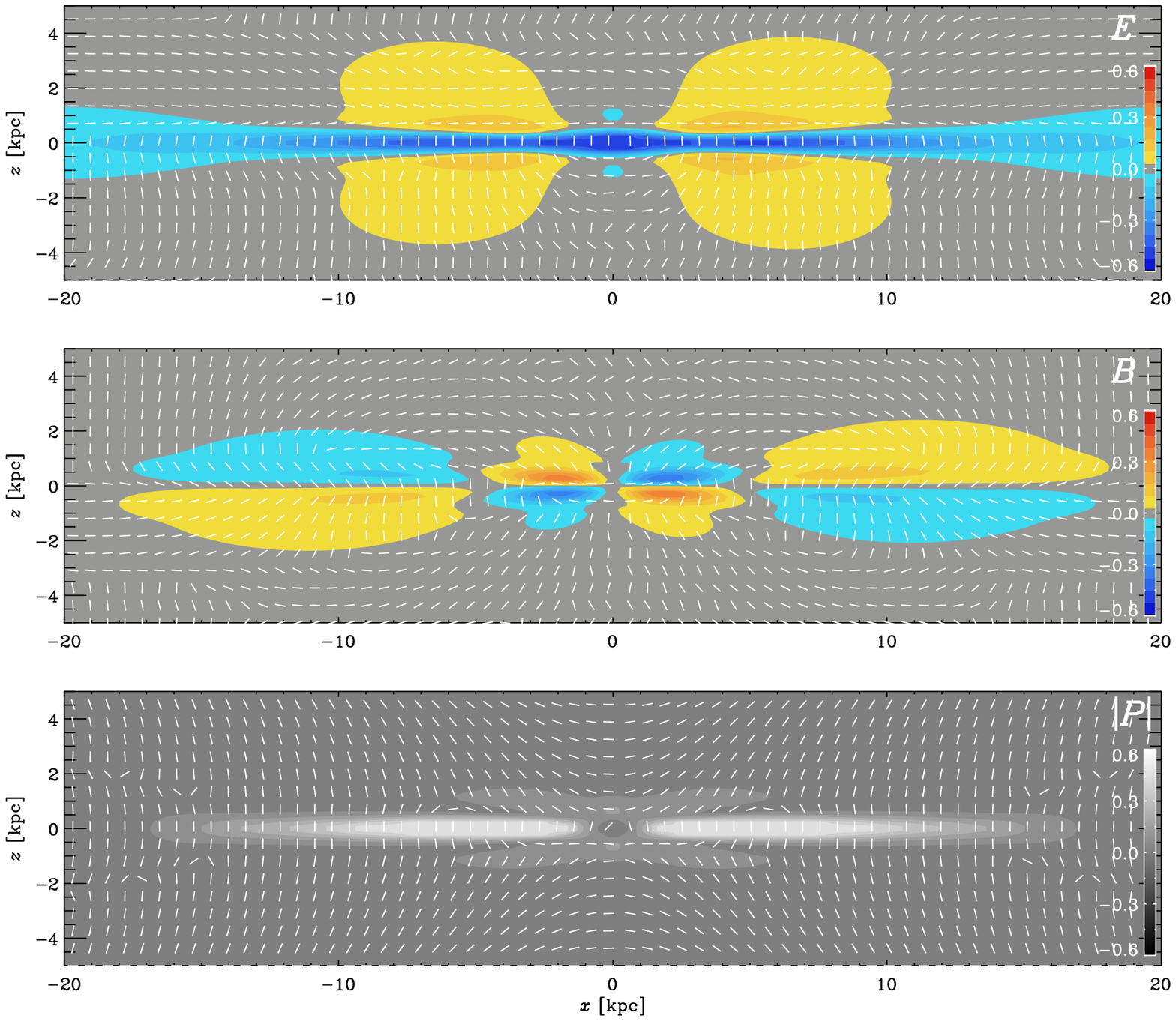}
\end{center}\caption[]{
Same as \Fig{pEBxz}, but for Model~III with
$W_r=10\kms$ and $H_\alpha=H_W=0.5\kpc$.
}\label{pEBxz_thick_pol_128x64_20kc_BDMSST93_wind10a}\end{figure*}

On the periphery of the computational domain, we assume normal-field
conditions, i.e., $\nnn\cdot\meanAA=(\nnn\cdot\nab)\times\meanAA=0$,
where $\nnn$ is the unit vector normal to the boundary.
The initial condition for $\meanAA(\xx,0)$ is Gaussian distributed
white noise of low amplitude.
In this paper, we measure lengths in $\!\kpc$ and speeds in
$\km\s^{-1}$, so time is measured in units of $0.98\Gyr$.
For simplicity, we drop the factor 0.98 when specifying times or inverse
times.

We consider several models.
First, we take the one of \cite{Bra15}, which was designed to reflect
typical spiral galaxies (Model~I).
Second, we modify the parameters (rotation curve and $\alpha$ effect)
such that they match those of the model of \cite{BDMSST93} of NGC~891
(Model~II).
In both cases, we use $64^2\times16$ mesh points, which proved sufficient
to resolve the spatial structure of the magnetic field.
Calculations with $128^2\times32$ mesh points produced virtually
identical polarization maps.
Next, we use a model with a wind and a thinner disk, where
the polarization pattern resembles more closely the observed ones (Model~III).
Finally, to study the properties of our diagnostic as a proxy for magnetic
helicity in terms of the $B$ polarization, we also present a model with
a negative $\alpha_0$, where we expect the sign of the magnetic helicity
to be reversed relative to what it is otherwise (Model~IV).
However, the nature of the $\alpha\Omega$ dynamo also changes and the
most easily excited field is then an oscillatory one, so no direct
comparison is possible.

The computations are performed with the {\sc Pencil Code} \citep{PC}
using either Cartesian coordinates (Models~I, II, and IV) or
cylindrical coordinates (Model~III).
We emphasize in this connection that we are dealing here with a mean
field model, so no sharp structures are expected to occur in such a case.
In \Tab{Tmodels} we summarize the various parameters used here.
We also give the maximum field strength $B_{\max}$.
The underlying simulation data are available online \citep{BF20}.

\begin{figure*}\begin{center}
\includegraphics[width=\textwidth]{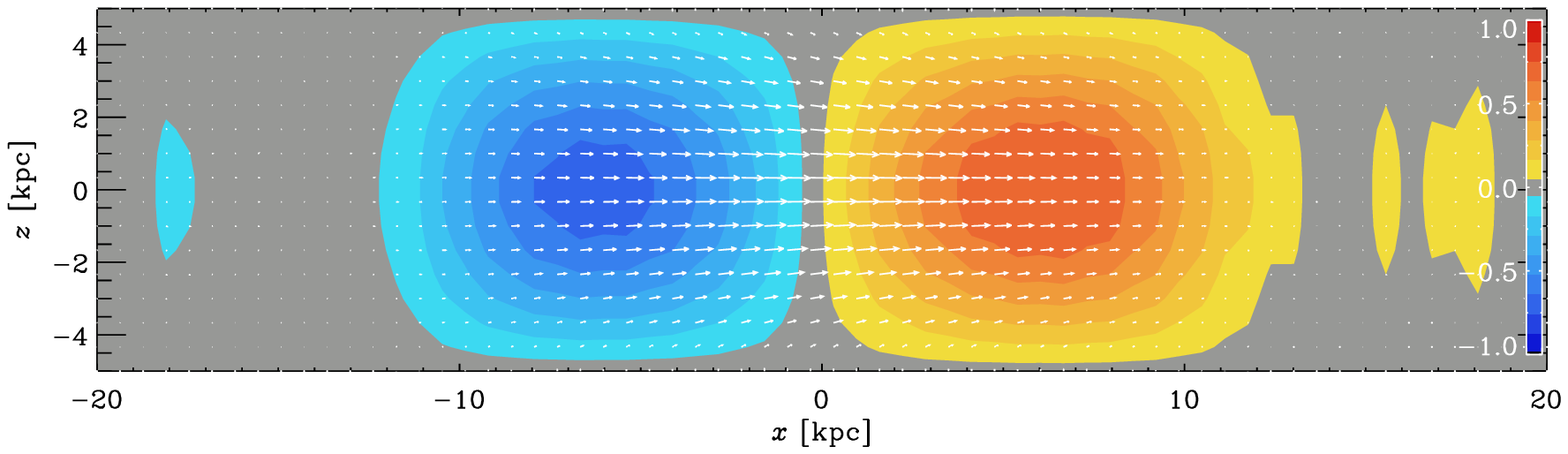}
\end{center}\caption[]{
Visualization of the magnetic field for Model~II showing
vectors of $\meanbb_\perp$ superimposed on a color representation of the
line-of-sight component in the same slice for which $E$ and $B$ are
shown in \Fig{pEBxz_64x64_15kc_BDMSST93c_Oml}.
Here, $b_z$ has been scaled by a factor of 20 relative to $b_x$, but
the vertical field still seems almost completely negligible.
}\label{pbfieldxz_64x64_15kc_BDMSST93c_Oml}\end{figure*}

\begin{table}\caption{
Parameters of the dynamo models discussed.
}\vspace{12pt}\begin{tabular}{lccccccc}
& $\Omega_0$ & $W_r$ & $\alpha_0\quad$ & $H_\alpha$ & $\varpi_\alpha$ & $\!\!\varpi_{\alpha\Omega}\!\!$ & $B_{\max}$\\
&$\!\![\!\Gyr^{-1}]\!\!\!$ & $\!\![\!\kms]\!\!$ & $\!\![\!\kms]\!\!$ & $\![\!\kpc]\!$ & $\!\![\!\kpc]\!\!$ & & $\!\![\!\uG]\!\!$\\
\hline
I   & $100$ &$\;0$ &$\;\;\,16$& $0.2$ & --- & $\infty$            & 8.7 \\
II  & $ 75$ &$\;0$ &$\;\;\,22$& $1.5$ & 15  & $\varpi_\Omega\!\!$ &15.6 \\
III & $ 75$ &$10\;$&$\;\;\,22$& $0.5$ & 15  & $\varpi_\Omega\!\!$ & 0.72 \\
IV  & $ 75$ &$\;0$ &$-15$     & $1.5$ & 15  & $\varpi_\Omega\!\!$ & 0.82 \\
\label{Tmodels}\end{tabular}\end{table}

\section{Results}

\subsection{$E$ and $B$ polarizations}

Our models lead to an early exponential growth of the magnetic field
and reach saturation after about $10\Gyr$.
Here we consider only the saturated phase of the models.
As a first step, we show in \Fig{pEBxz} the $E$ and $B$ polarizations
together with the polarized intensity and polarization orientations
for Model~I from the $xz$ plane near $y=-10\kpc$.
In \Fig{pEBxz_64x64_15kc_BDMSST93c_Oml} we again show the $E$ and $B$
polarizations, but now for Model~II.
The polarized intensity and polarization orientations show just a
horizontal pattern.
It is therefore amazing that the decomposition into its rotationally
invariant constituents appears so much richer.
The primary reason for this is connected just with the spatial variation
of the polarized intensity.
This is demonstrated in \App{LocGlob}, where we show the $E$ and $B$
polarizations for a polarization signal where Stokes $U=0$ and Stokes $Q$
consists of just a Gaussian.

There are remarkable similarities between the polarization patterns
obtained from a Gaussian and those obtained from our dynamo model.
This suggests that the information content in the overall pattern seen
in \Figs{pEBxz}{pEBxz_64x64_15kc_BDMSST93c_Oml} is not very profound,
although more physically meaningful information may still be hidden
within the more subtle departures from this overall pattern.

The pixel resolution in \Fig{pEBxz_64x64_15kc_BDMSST93c_Oml} is about
$650\pc$, which is the spatial resolution when NGC~891 is observed at a
distance of $9.6\Mpc$ \citep{Strickland} with the 14 arcsecond beam of
the JCMT at $850\um$ wavelength.
We see that, in both cases, $E$ is symmetric about the midplane $z=0$,
and $B$ is antisymmetric.
Moreover, $E$ is negative in the halo near the axis ($x=0$) and positive
near the midplane further away from the axis.
In Model~I, however, $E$ is positive at the center ($x=z=0$), while in
Model~II it is negative.
As expected, $B$ changes sign about the equator.
It also changes sign between the two sides of the rotation axis.
In the first and third quadrants, the sign is negative, while in the
second and fourth, it is positive.
The arrow-less vectors associated with the $B$ polarization
show a clockwise inward swirl in the first and third quadrants and an
anti-clockwise inward swirl in the second and fourth quadrants.
This signature seems to be surprisingly independent of the differences
between Models~I and II.
The patterns of $E$ and $B$ are also rather smooth; the typical scale
would be the scale height of the magnetic field, which is about $2\kpc$.
In this context, we must recall that our model can only deliver the mean
magnetic field, and that the actual magnetic field must also contain a
fluctuating component of smaller scales below $100\pc$.

To get a sense of the range of $E$ and $B$ polarization patterns, we
show in \Fig{pEBxz_thick_pol_128x64_20kc_BDMSST93_wind10a} the results
for a model with a wind of just $10\kms$ and a thinner disk;
see Model~III in \Tab{Tmodels}.
Here we inspect the plane $y=0$, where the polarization
orientations best resembles those observed in synchrotron emission
\citep{HS99,Wie+15,Kra19,Kra+20}.
The maximum field strength is here much weaker.
This can be explained by the wind advecting the magnetic field
away from the disk.
In our models, this suppression of the magnetic field might be
more effective because of the anisotropic $\alpha$ effect.
Interestingly, the $E$ polarization is now strongly negative along
the midplane.
This is similar to what is seen for our Galaxy viewed from the position
of the Sun (BB20).
Furthermore, the polarization is now more complicated.
The patches with negative (positive) $B$ values in the first and third
(second and fourth) quadrants are now closer to the axis, and there
is an additional such pattern with opposite signs further out.

\subsection{Magnetic field and helicity}

It is important to realize that neither the helicity of the magnetic
field nor any proxy of it can straightforwardly be extracted from just
a simple inspection of the morphology of the magnetic field.
This is mostly because of the strong dominance of the azimuthal magnetic
field over the vertical ($z$) field in galactic latitude.
This becomes clear once again from
\Fig{pbfieldxz_64x64_15kc_BDMSST93c_Oml}, where we show vectors of
$\meanbb_\perp$ superimposed on a color scale representation of $b_y$.
Here, $b_z$ has been scaled by a factor of 20 relative to $b_x$, but
the vertical field still seems almost completely negligible.

It is long known that the magnetic field in galaxies tends to have
even symmetry about the equatorial plane \citep{BBMSS96,BCEB19}.
This is also the case for all of our models.
This means that the azimuthal magnetic field is symmetric about the
midplane and the vertical magnetic field is antisymmetric about in
the midplane, just as seen from \Fig{pbfieldxz_64x64_15kc_BDMSST93c_Oml}.
Note, also, that the strongest magnetic field is found at a distance of
about $7\kpc$ from the center.
The results for Model~I are similar, except that here the strongest
magnetic field occurs at a distance of about $5\kpc$ from the center.

\begin{figure}\begin{center}
\includegraphics[width=\columnwidth]{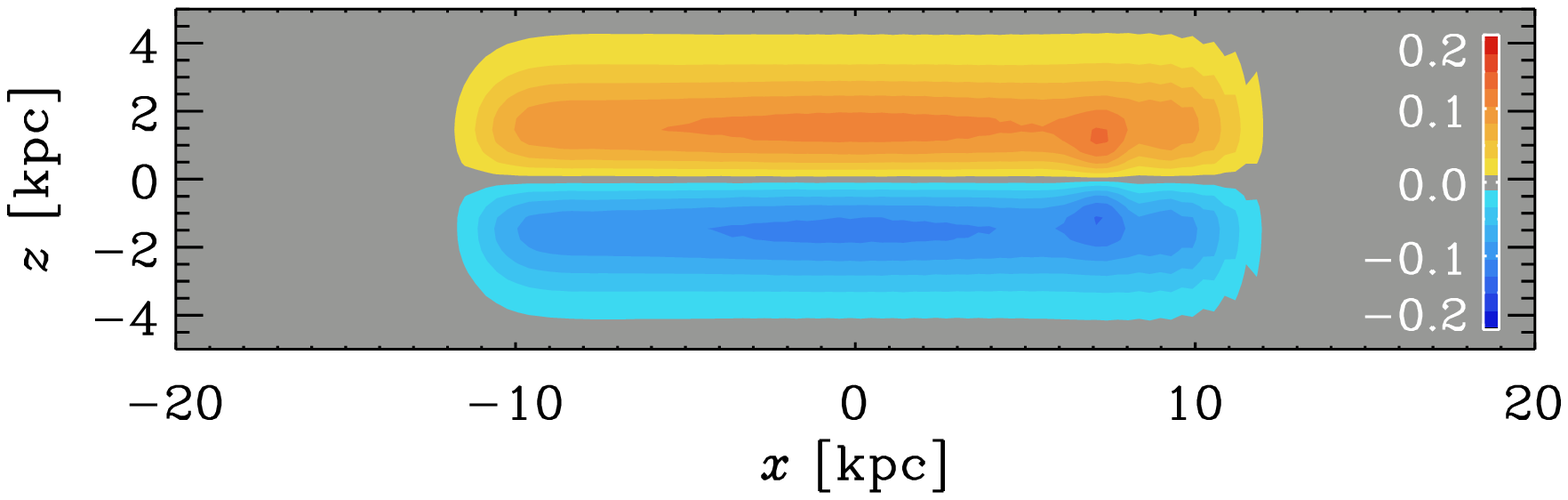}
\end{center}\caption[]{
Visualization of the normalized current helicity density,
$\mu_0\meanJJ\cdot\meanBB$, for Model~II at $y=-10\kpc$.
}\label{pphelicityxz_128x128x32_20kc_BDMSST93c_Oml}\end{figure}

\begin{figure}\begin{center}
\includegraphics[width=\columnwidth]{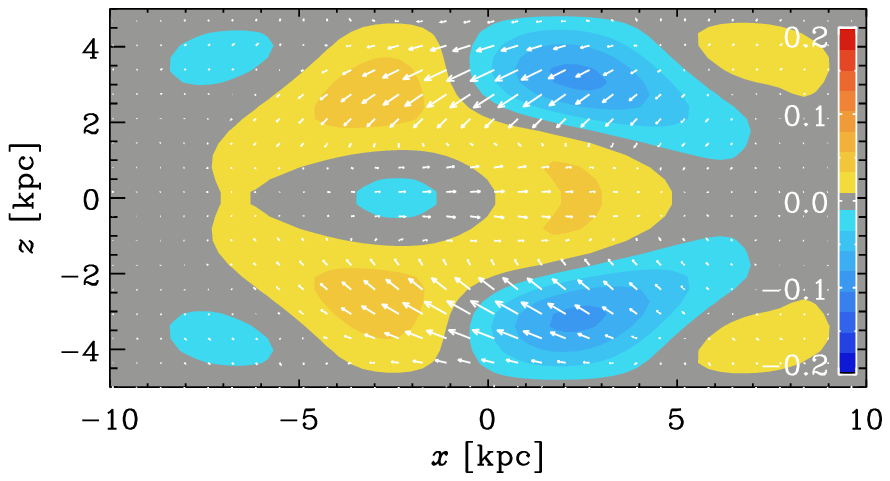}
\end{center}\caption[]{
Visualization of the magnetic field showing vectors of $\meanbb_\perp$
superimposed on a color representation of the line-of-sight component,
similar to \Fig{pbfieldxz_64x64_15kc_BDMSST93c_Oml}, but for Model~IV, at
$y=-10\kpc$, and $b_z$ has been scaled by a factor of 5 relative to $b_x$.
The plot range in the $x$ direction has been clipped to $\pm10\kpc$.
}\label{pbfieldxz_128x128x32_20kc_BDMSST93c_Oml_neg5}\end{figure}

To make contact with the helicity of the magnetic field in our
model, we plot in \Fig{pphelicityxz_128x128x32_20kc_BDMSST93c_Oml} a vertical
slice of the current helicity density $\meanJJ\cdot\meanBB$.
We can see that $\meanJJ\cdot\meanBB$ is mostly
positive (negative) in the upper (lower) disc plane.
This agrees with the sign of $\alpha_\perp$, which is also positive
(negative) in the upper (lower) disc plane.
For $Q_\alpha=25$, the normalized current helicity density,
$\mu_0\meanJJ\cdot\meanBB$, would be in units of $\uG^2\kpc^{-1}$.

\begin{figure}\begin{center}
\includegraphics[width=\columnwidth]{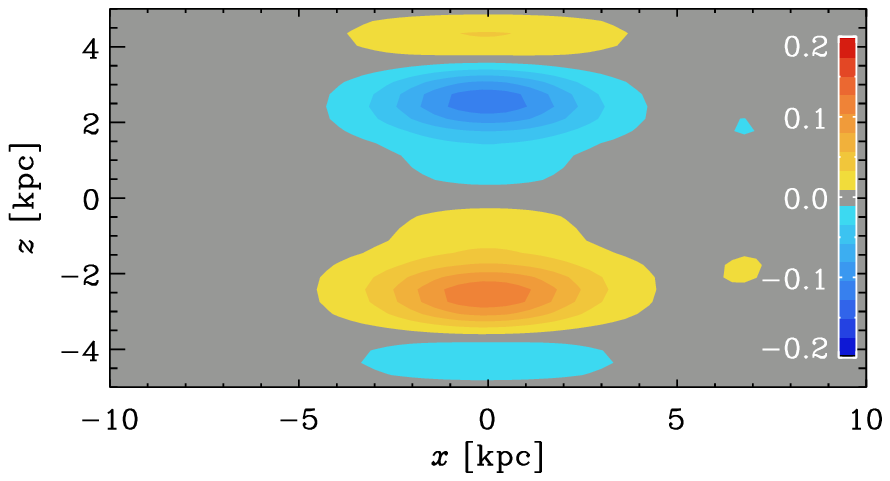}
\end{center}\caption[]{
$\mu_0\meanJJ\cdot\meanBB$, similar to
\Fig{pphelicityxz_128x128x32_20kc_BDMSST93c_Oml},
but for Model~IV, and again at $y=-10\kpc$.
}\label{pphelicityxz_128x128x32_20kc_BDMSST93c_Oml_neg5}\end{figure}

\begin{figure}\begin{center}
\includegraphics[width=\columnwidth]{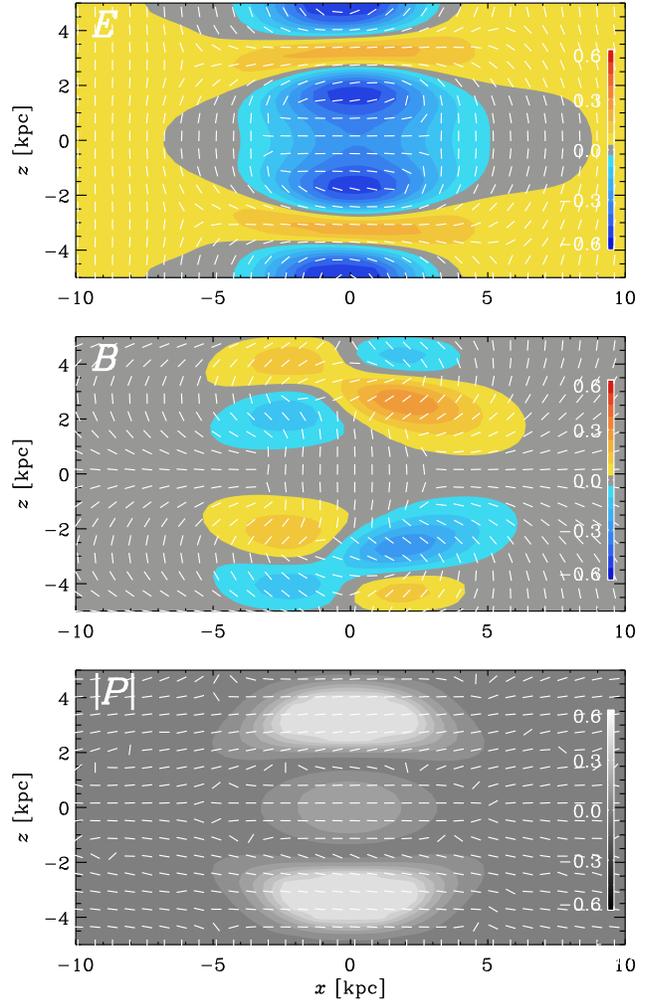}
\end{center}\caption[]{
Same as \Fig{pEBxz}, but for Model~IV with negative $\alpha_0$ at an
arbitrarily chosen time.
The plot range in the $x$ direction has been clipped to $\pm10\kpc$.
}\label{pEBxz_128x128x32_20kc_BDMSST93c_Oml_neg5}\end{figure}

\subsection{Negative $\alpha$ and reversed helicity}

To give a sense of the range of possibilities for the $E$ and $B$
patterns in models of edge-on galaxies, we now consider Model~IV with
a negative $\alpha_0$.
Physically, negative values of $\alpha_0$ can be caused by several
different mechanisms.
One such mechanism is given when magnetic driving dominates over kinetic
driving, in particular when magnetic buoyancy dominates over thermal
buoyancy.
In such cases, it has been found that $\alpha_0$ may be negative
\citep{BNST95,RP00}.
This is the case when turbulence is driven by the magneto-rotational
instability, which can play a role also in galaxies \citep{MNKASM13}.

As mentioned in \Sec{TheModel}, a model with a negative value
of $\alpha_0$ is not simply a mirror image of a model with a positive
value of $\alpha_0$, because the nature of the dynamo also changes.
In the present case, the field becomes oscillatory, but retains
its even symmetry with respect to the midplane; see
\Fig{pbfieldxz_128x128x32_20kc_BDMSST93c_Oml_neg5} for a visualization
of the magnetic field in a cross-section through $y=-10\kpc$.
Note that the magnetic field is fairly much confined within a cylindrical
radius of $10\kpc$.
This is why we have clipped in
\Fig{pbfieldxz_128x128x32_20kc_BDMSST93c_Oml_neg5} (and the following
two figures) the $x$ range beyond $\pm10\kpc$.
However, the numerical calculations have been carried out, just as before,
in the larger domain with $|x|,|y|\leq20\kpc$.

To verify that the magnetic helicity has indeed changed sign, we show
in \Fig{pphelicityxz_128x128x32_20kc_BDMSST93c_Oml_neg5} the current
helicity, again through the plane $y=-10\kpc$.
We see that $\meanJJ\cdot\meanBB$ is now indeed negative in the upper
disc plane, but there is an additional sign change at
large heights above and below the plane of a galaxy.

Finally, the resulting $E$ and $B$ polarizations at $y=-10\kpc$
are shown in \Fig{pEBxz_128x128x32_20kc_BDMSST93c_Oml_neg5}.
Note that $B$ has changed sign with respect to
\Figs{pEBxz}{pEBxz_64x64_15kc_BDMSST93c_Oml}, at least
sufficiently close to the midplane.
Closer to the boundaries at $z=\pm5\kpc$, this sign has
changed again, somewhat similarly to what was seen in
\Fig{pphelicityxz_128x128x32_20kc_BDMSST93c_Oml_neg5} for
$\meanJJ\cdot\meanBB$.
The $E$ polarization is somewhat different than before, although the
basic features are unchanged: it is negative near the axis and positive
further away from it, except right at the midplane.
Within the midplane, the negative sign of $E$ near the axis extends now
also to larger radii.

Unlike \Fig{pEBxz_64x64_15kc_BDMSST93c_Oml}, the polarized intensity
now shows concentrations a certain distance away from the midplane.
The polarization orientations are still mostly horizontal, so the basic
phenomenology explained in \App{LocGlob} still applies.
This explains then why the $E$ and $B$ polarizations in
\Fig{pphelicityxz_128x128x32_20kc_BDMSST93c_Oml_neg5} show a doubling
of the features seen in \Fig{pEBxz_64x64_15kc_BDMSST93c_Oml}.

\section{Discussion and future prospects}

Edge-on galaxies provide an opportunity to study some basic
aspects of magnetic fields in galaxies.
This can have implications for the dynamo interpretation of their
generation.
Our preliminary investigation based on simple models suggests that the $E$
polarization is positive near the disc midplane and away from the axis,
where it tends to outline star-like patterns in the magnetic field, while
in the halo near the axis it is negative, corresponding to ring-like
patterns in the magnetic field; see the top panel of \Fig{pEBxz}.
The $B$ polarization is negative in the first and third quadrants and
outlines counterclockwise inward spiraling patterns, while in the second
and fourth quadrants, we have positive values, corresponding to clockwise
inward spiraling patterns; see the bottom panel of \Fig{pEBxz}.
Again, we emphasize here that our discussion of spiraling polarization
patterns concerns the edge-on view of the $B$ polarization and is not
connected with the spiraling appearance of galaxies viewed face-on.

Our hope is that the predictions for the signs of the $E$ and $B$
polarizations could soon be verified observationally.
In addition to utilizing already existing synchrotron polarimetry data, we
discuss here the possibility of using observations of dust polarization.
Dust polarization observations have proven to be an excellent tool to
measure the orientation of the magnetic field in star-forming
regions and to assess the relative magnitudes of the mean and turbulent
components of the field \citep{Hildebrand}.
Linear polarization imaging provide the orientation of the plane-of-sky
magnetic field lines, as non-spherical dust grains tend to align with
their short axis perpendicular to the direction of the magnetic field,
giving rise to linear polarization of the emission.
As discussed by \cite{Andersson}, the radiative alignment torque (RAT)
theory is the most widely accepted mechanism for interstellar dust grains
to align in a realistic environment; see also \cite{Bracco19b}.
All-sky dust polarization imaging by {\em Planck} at $850\um$ have
already revealed well-defined magnetic field patterns in the ISM from
Galactic scales down to molecular cloud scales of $\ga10\pc$.
Because there are no observational studies to test such a mechanism toward a
galaxy with a spatial resolution of $10^2$--$10^3\pc$, it is important
to assess the validity of such a mechanism.

Although a size of $\ga10\pc$ is still one or two orders of magnitude
larger than that of dense cloud cores where stars and stellar cluster
form, it is worth attempting to investigate the magnetic-field structure
to study the physical conditions of $\la10\pc$-scale molecular
cloud formation.
Submillimeter (submm) emission polarimetry is sensitive to tracing the column
density of the cold ($T\la$ a few times $10\K$) ISM, whereas near-infrared
(NIR) absorption polarimetry may have suffered from scattering, while
cm-wavelength radio polarimetry may already begin to be affected by Faraday rotation.
Indeed, recent polarization observations with SOFIA HAWC+ toward the
archetypical starburst galaxy M\,82 at 53 and $154\um$ \citep{Jones} have
clearly demonstrated that submm/FIR emission polarimetry offers another
way of probing the magnetic field structure by observing entrained dust
grains by the super galactic wind.
The authors did not identify any component expected from the
large-scale dynamo field.
Nevertheless, as the next step, it is worth considering longer wavelength
polarimetry at $850\um$ for large heights above and below the plane of a galaxy
to unveil the magnetic-field structure in the cold ISM of that galaxy.
This would give us a more specific hint at how the magnetic field in a
galaxy is maintained against diffusion by turbulence.
Moreover, we point out that previous Herschel PACS and SPIRE imaging at $100$,
$160$, and $250\um$ revealed that NGC~891 is rich in dust grains---even
toward its halo; the scale height of the dusty halo is $1.44\pm0.12\kpc$
\citep{Bocchio}.
These authors showed that $\sim2$--$3\%$ of the mass of the dust is
present further than $2\kpc$ from the midplane.
This agrees with the analysis from the $450$ and $850\um$ emission taken with
SCUBA \citep{Alton+98}.
We therefore argue that the method proposed in this work may be feasible;
it is complementary to the conventional two methods (radio synchrotron
and NIR polarimetry).

At wavelengths of $850\um$, the emission is optically thin as shown
by the graybody-fitting at the submm/FIR spectral energy distribution
(SED); see Fig.~3 in \cite{Alton+98}.
Their SED analysis clearly favors a frequency index of grain emissivity,
$\beta$, of 2 rather than 1.5, suggesting that grains' properties are
ISM-like, i.e., not evolved like those seen in protoplanetary disks.
These results were later confirmed by multi-wavelength imaging studies
adding Spitzer, WISE, and Herschel data \citep{WIMGB09,Hughes}.
The grain emissivity, its frequency index, and the dust
temperature are similar to those measured in $\ga10\pc$-scale galactic
molecular clouds where submm polarization imaging were performed
\citep{Matthews+09,Ward-Thompson+17}.
However, caution must be exercised because polarization properties of grains
on galaxy scales are poorly known.
Nevertheless, given the highly-enhanced sensitivity of the
current instrument, SCUBA-2 plus the POL-2 polarimeter system of
$\sim4\mJy\beam^{-1}$ in polarization intensity for a typical observation,
we do believe that $850\um$ polarimetric observations toward NGC\,891
could help to assess signs of handedness.
As of today, no polarization data in FIR or submm are yet available
for NGC~891.
However, intensity maps of NGC~891 at $850\um$ show two ``blobs'' or ``knots''
at a distance of about $4$--$5\kpc$ from the center \citep{HKB02,WIMGB09}.
This could also be suggestive of a ring-like magnetic field, similarly
to what is seen in \Fig{pbfieldxz_64x64_15kc_BDMSST93c_Oml}.
This would be analogous to the ring-like magnetic field of M~31
\citep{BBGM20}.
Interestingly, the ring-like concentration cannot easily be explained with
kinematic theory \citep{RSS88}, unless one invokes a similar variation
of the gas density and thereby of $\Beq$ \citep{BBMSS96}.
Such details have been ignored in our present modeling.

A potential shortcoming of the present calculations is that our
predictions are based solely on mean-field dynamos.
This means that we only take the large-scale component of the magnetic
field into account.
In reality, there is also a small-scale component whose helicity is
expected to have the opposite sign \citep{BB03}.
How this affects the detection prospects of systematic $E$ and $B$
polarizations is currently unknown and should be a target of future
investigations.
The typical scale of fluctuations would not exceed $100\pc$ \citep{BBMSS96},
which is well below the resolution scale of the James Clerk Maxwell Telescope.
It is therefore possible that our present results may already give a
good hint at what can be observed in the near future.
Higher resolution observations would be of great interest for testing
the idea of a magnetic helicity reversal at smaller length scales.
In the meantime, however, it would be worthwhile to explore this regime
of smaller length scales with detailed turbulence simulations without
adopting mean-field theory.

Another opportunity for improvements is given by our increased knowledge
of the rotation curves of NGC~891 \citep{FSK11}.
In particular, it is now known that the rotation in the halo of NGC~891
is slower than that in the galactic disc \citep{OFS07}.
This implies the presence of vertical shear that could modify the vertical
helicity profile of the galaxy.

\section*{Acknowledgements}

This work was supported in part through the Swedish Research Council,
grant 2019-04234, and the National Science Foundation under the grant
AAG-1615100.
This research is partially supported by Grants-in-Aid for Scientific
Researches from the Japan Society for Promotion of Science (KAKENHI
19H0193810).
We acknowledge the allocation of computing resources provided by the
Swedish National Allocations Committee at the Center for Parallel
Computers at the Royal Institute of Technology in Stockholm.
\vspace{2mm}

\noindent
{\em Code and data availability.} The source code used for the
simulations of this study, the {\sc Pencil Code} \citep{PC},
is freely available on \url{https://github.com/pencil-code/}.
The DOI of the code is https://doi.org/10.5281/zenodo.2315093.
The simulation setup and the corresponding data are freely available on
\url{https://doi.org/10.5281/zenodo.3897954}.

\appendix
\section{$E$ and $B$ from a patch in $Q$}
\label{LocGlob}

\begin{figure}\begin{center}
\includegraphics[width=\columnwidth]{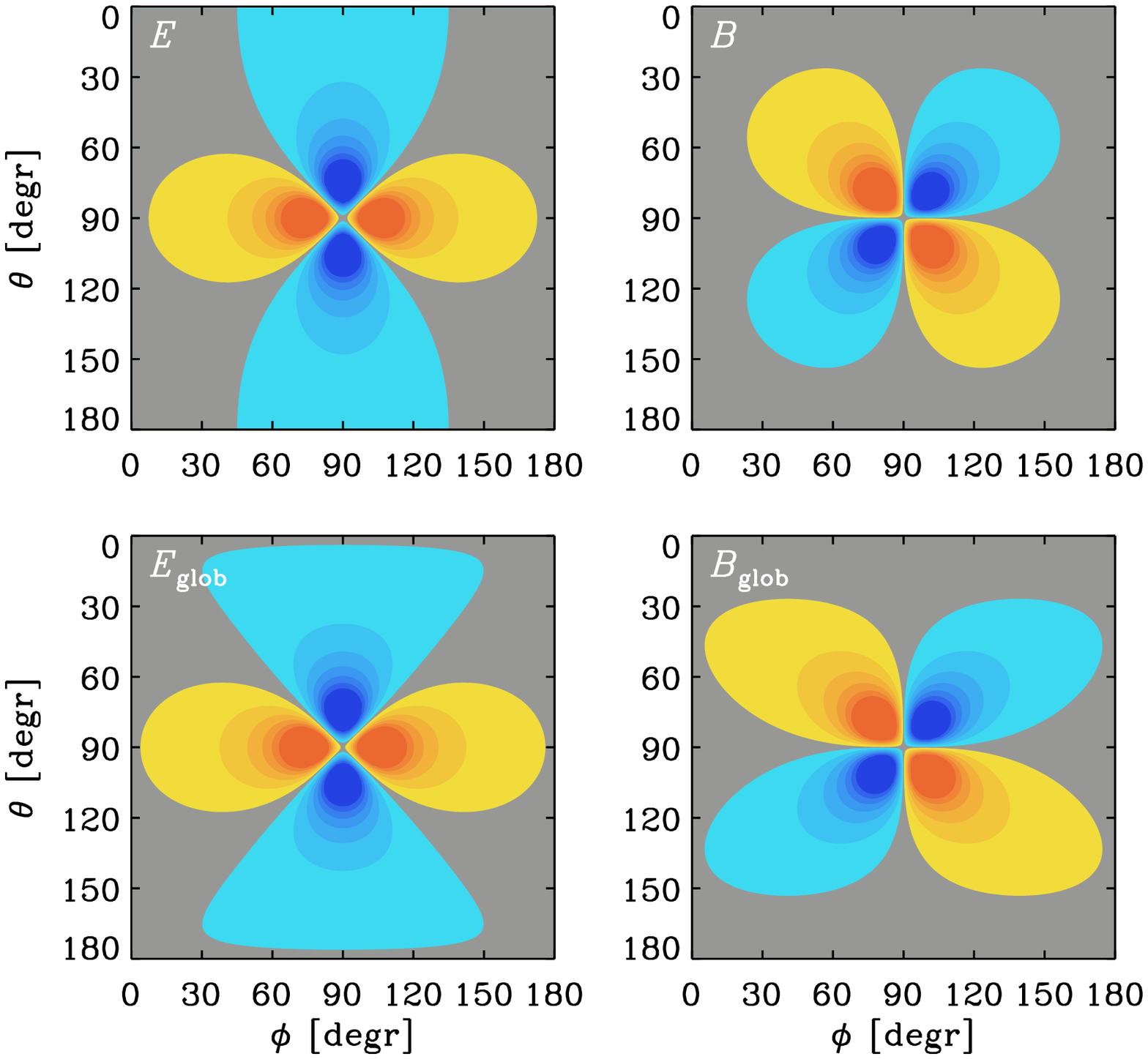}
\end{center}\caption[]{
Local representations $E$ and $B$ together with their global
counterparts $E_{\rm glob}$ and $B_{\rm glob}$ for a Gaussian
patch with just a horizontal magnetic field, $\meanbb=(0,\meanb_\phi)$.
}\label{pcomp2d}\end{figure}

The purpose of this appendix is to compute the $E$ and $B$ polarization
patterns for a signal with Stokes $U=0$ and Stokes $Q$ originating from
just a Gaussian patch of a projected horizontal or a vertical magnetic
field.
To also illuminate the question about different sign conventions, we use
this opportunity to compare the local approach given by \Eqs{Rk}{EBdef}
with the global one, where the complex polarization is expanded in terms
of spin-2 spherical harmonic functions \citep[e.g.,][]{Dur08}.
We therefore work with spherical coordinates $(\theta,\phi)$, where
$\theta$ is colatitude and $\phi$ is longitude with $0<\theta<\pi$
and $0<\phi<2\pi$.
We write the projected magnetic field in the form
\begin{equation}
\meanb_{\phi/\theta}=b_0\exp\left\{-\left[(\theta-\theta_0)^2+(\phi-\phi_0)^2\right]/
2\sigma^2\right\},
\label{QGauss}
\end{equation}
where we choose $\theta_0=\phi_0=\pi/2$ to be a point on the equator
and $\sigma=10\degr$ is chosen for the size of the patch.
(The underlying three-dimensional magnetic field would be always solenoidal.)
The global representations of $E$ and $B$ (indicated by subscripts `glob')
are expressed through the complex function
\EQ
E_{\rm glob}+\ii B_{\rm glob}\equiv R=\sum_{\ell=2}^{N_\ell}\sum_{m=-\ell}^{\ell}
\tilde{R}_{\ell m} Y_{\ell m}(\theta,\phi),
\label{EBfromQU}
\EN
where the coefficients $\tilde{R}_{\ell m}$ are given by
\EQ
\tilde{R}_{\ell m}=\int_{4\pi}
(Q+\ii U)\,_2 Y_{\ell m}^\ast(\theta,\phi)\,
\sin\theta\,\dd\theta\,\dd\phi.
\label{QUfromEB}
\EN
Here $_2 Y_{\ell m}^\ast(\theta,\phi)$ are the spin-2 spherical
harmonics of spherical harmonic degree $\ell$ and order $m$ and
the asterisk denotes complex conjugation.
In the cosmological context, both $E_{\rm glob}$ and $B_{\rm glob}$
are defined with a minus sign; see, e.g., Eq.~(6) of \cite{ZS97}.
The results for a Gaussian $\meanb_\phi$ and a $\meanb_\theta$ patch given by
\Eq{QGauss} are shown in \Figs{pcomp2d}{pcomp2d_vert}, using
$N_\ell=36$ as the truncation level.
We compare $E_{\rm glob}$ and $B_{\rm glob}$ with $E$ and $B$ in the
range $0<\phi<\pi$, but the calculations have been done in the full
range $0<\phi<2\pi$.

\begin{figure}\begin{center}
\includegraphics[width=\columnwidth]{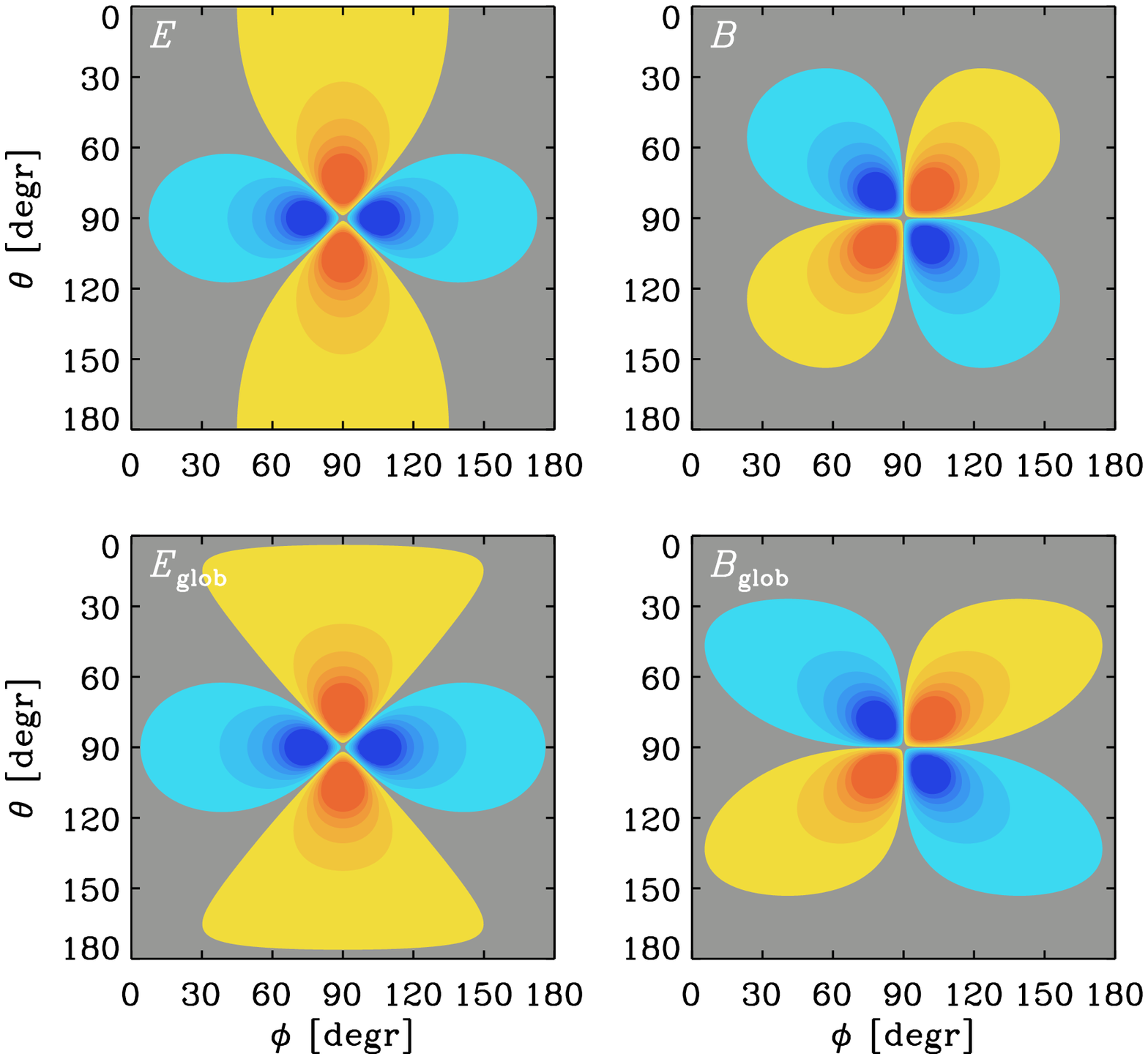}
\end{center}\caption[]{
Same as \Fig{pcomp2d}, but for a vertical magnetic field, $\meanbb=(\meanb_\theta,0)$.
}\label{pcomp2d_vert}\end{figure}

Both $E$ and $E_{\rm glob}$ as well as $B$ and $B_{\rm glob}$ agree with
each other and are qualitatively similar to the overall appearance of $E$
and $B$ seen in \Fig{pEBxz_64x64_15kc_BDMSST93c_Oml}.
The agreement between local and global representations shows that for
a single patch, the sign convention used in \cite{Dur08} and
\cite{Bra19} for the global representation agrees with the local one
used in BBKMRPS and \cite{Pra+20}.

Interestingly, the cloverleaf-shaped pattern of $B$ in \Fig{pcomp2d}
resembles a similar pattern found by BB20 for the Galactic center.
A difference, however, is the sign of both $E$ and $B$.
In \Fig{pcomp2d_vert} we show that such a pattern can explained be
by a case with a strong vertical magnetic field.
As expected, the signs of both $E$ and $B$ are now changed and
the resulting pattern matches that found BB20.


\end{document}